%% file: 00_main.tex
\documentclass[journal,comsoc]{IEEEtran}
\ifCLASSINFOpdf
\else
\fi
\usepackage[cmex10]{amsmath}
\usepackage{amssymb}
\usepackage{siunitx}
\usepackage{units}
\usepackage{caption}
\usepackage{subcaption}
\usepackage{algpseudocode}
\usepackage{algorithm}
\algnewcommand{\Initialization}[1]{%
  \State \textbf{initialization:}
  \Statex \hspace*{\algorithmicindent}\parbox[t]{.8\linewidth}{\raggedright #1}
}

\algnewcommand{\Rep}[1]{%
  \State \textbf{repeat:}
  \Statex \hspace*{\algorithmicindent}\parbox[t]{.8\linewidth}{\raggedright #1}
}

\usepackage{tikz}
\usetikzlibrary{shapes,arrows,fit,positioning,calc}
\usetikzlibrary{plotmarks}
\usetikzlibrary{decorations.pathreplacing}
\usetikzlibrary{patterns}
\usetikzlibrary{arrows,automata}
\usetikzlibrary{spy}
\usepackage{graphicx}

\usepackage{pgfplots}
\pgfplotsset{compat=newest}

\newcommand{\pr}[1]{\ensuremath{\left[#1\right]}}
\newcommand{\pc}[1]{\ensuremath{\left(#1\right)}}
\newcommand{\chav}[1]{\ensuremath{\left\{#1\right\}}}
\newcommand{\PM}[1]{\ensuremath{\left|#1\right|}}

\definecolor{r}{rgb}{1, 0, 0}

\definecolor{dark_green}{rgb}{0, 0.33, 0.13}
\definecolor{naplesyellow}{rgb}{0.99, 0.93, 0.0}
\definecolor{aureolin}{rgb}{1, 0.8, 0}
\definecolor{purple}{rgb}{0.4940 0.1840 0.5560}

\begin{document}
%
\title{MMSE Symbol Level Precoding Under a Per Antenna Power Constraint for Multiuser MIMO Systems With PSK Modulation}

\author{Erico~S.~P.~Lopes,~\IEEEmembership{Graduate Student Member,~IEEE} and~Lukas~T.~N.~Landau,~\IEEEmembership{Senior Member,~IEEE,}\vspace{-1em}				
\thanks{The authors are with Centro de Estudos em Telecomunica\c{c}\~{o}es, Pontif\'{i}cia Universidade Cat\'{o}lica do Rio de Janeiro, Rio de Janeiro CEP 22453-900, Brazil, (email: \{erico, lukas.landau\}@cetuc.puc-rio.br).} 
\thanks{This work has been supported by the {ELIOT ANR18-CE40-0030 and FAPESP 2018/12579-7} project.}
}
\maketitle


\begin{abstract}
This study proposes a symbol-level precoding algorithm based on the minimum mean squared error design objective under a strict per antenna power constraint for PSK modulation. The proposed design is then formulated in the standard form of a second-order cone program, allowing for an optimal solution via the interior point method. Numerical results indicate that the proposed design is superior to the existing approaches in terms of bit-error-rate for the low and intermediate SNR regime.
\end{abstract}
\begin{IEEEkeywords}
MMSE, symbol-level precoding, strict per antenna power constraint, MIMO systems, second-order cone programming.
\end{IEEEkeywords}



%

\input{01_Introduction.tex}

\input{02_System_Model.tex}
\input{03_State_of_the_art.tex}
\input{04_Precoding_task.tex}
\input{05_Numerical_Results.tex}
\input{06_Conclusions.tex}

\bibliographystyle{IEEEtran}
\bibliography{bib-refs}

\end{document}

%% file: 01_Introduction.tex
\section{Introduction}

Massive multiuser multiple-input multiple-output (MU-MIMO) systems are considered as a promising technology and are expected to be vital for future wireless communications networks \cite{6G_Future_Directions}. For MU-MIMO systems a fundamental problem is the design of low-complexity precoding algorithms that attain the high reliability constraints of future wireless communications networks.

Linear techniques such as zero forcing (ZF) and matched filtering \cite{Kammoun2014,Tenbrink_2013} are known to be asymptotically optimal \cite{Power_consumption} for massive MIMO systems due to the favorable propagation effects that rise for infinitely large arrays. However, when considering linear precoding, an established assumption in the literature \cite{M_Joham_ZF,Swindlehurst2005} is that the transmit symbols are constrained by an average total power constraint (TPC). This yields a system that is easier to model, yet, according to \cite{CVX-CIO}, in a realistic scenario each base station (BS) antenna is connected to its own power amplifier (PA) and thus has to meet its specific power constraints. 

With this, several precoding techniques arose considering per antenna power constraints (PAPC). Linear channel-level precoding strategies considering an average PAPC are well studied in the literature \cite{WeiYu2007,BoccardiSPAWC2006,zf_papc,MRT_papc}. However, according to \cite{lmmse_papc}, the consideration of a strict PAPC (SPAPC) yields a more realistic scenario since the transmit power at each antenna is upper bounded by a threshold to avoid sever distortion at the PA due to clipping. With this, different linear precoding techniques have been developed considering SPAPCs \cite{lmmse_papc,Pi2012}. More recently, the symbol-level precoding (SLP) strategy has been receiving increasing attention since it allows for a higher degree of reliability. In \cite{bjorn_globecom2016} a SLP method is devised considering a per antenna transmit power minimization under the condition of attaining QoS constraints for M-PSK modulations. In \cite{Bjorn_satellite} a SLP design is considered also for the minimization of the per antenna transmit power under QoS constraints in the context of satellite communications. 

Besides the aforementioned concept, two other design objectives have become prominent in the literature of SLP, the minimum mean squared error (MMSE) and maximum minimum distance to the decision threshold (MMDDT). The MMDDT objective (also known as constructive interference (CI) \cite{CVX-CIO} or maximum safety margin \cite{MSM_precoder}) is used considering a channel-level linear approach with a TPC in \cite{masouros_twc2018} and is well-established in the context of constant envelope precoding \cite{CVX-CIO, manifold_ce} and precoding under coarse quantization \cite{MSM_precoder,Landau2017,General_MMDDT_BB}. When considering the SPAPC the work from \cite{Chen2020} proposes two novel strategies based on the concept of \textit{strict} and \textit{non strict rotation} for CI-based precoding, where the \textit{non strict rotation} criterion is equivalent to MMDDT.

The MMSE utilization ranges from the established channel-level linear precoding strategy presented in \cite{M_Joham_ZF} to a SLP design considering coarse quantization \cite{lopes2021discrete}. Although prominent in the literature to the best of the authors' knowledge the MMSE objective has not been considered for SLP under a SPAPC.

In this study, we consider a SPAPC and propose a SLP algorithm based on the MMSE design criterion for PSK modulation. The proposed approach is formulated in the standard second-order cone programming (SOCPs) form which is readily solved with polynomial complexity using the interior points method (IPM). Numerical results indicate that the proposed method is superior to the existing techniques in terms of BER for the low and medium SNR regime.

The remainder of this paper is organized as follows: Section~\ref{sec:system_model} describes the system model. Section~\ref{sec:state_of_the_art} revises the state-of-the-art designs of SLP under a SPAPC. Section~\ref{sec:precoding} exposes the MMSE optimization problem, formulates it as a SOCP and provides the complexity analysis of the proposed algorithm. Section~\ref{sec:numerical_results} presents and discusses numerical results, while Section \ref{sec:conclusions} gives the conclusions.

Regarding the notation, bold lower case and upper case letters indicate vectors and matrices, respectively. Non-bold letters express scalars. The operators $(\cdot)^*$ and $(\cdot)^T$ denote complex conjugation and transposition, respectively. $S_{+}^{n}$ denotes the set of symmetric positive semidefinite matrices of dimension $n \times n$. Real and imaginary part operators are also applied to vectors and matrices, e.g., $\mathrm{Re}\left\{ \boldsymbol{x} \right\} = \left[ \mathrm{Re}\left\{ \left[\boldsymbol{x}\right]_1 \right\},\ldots,    \mathrm{Re}\left\{ \left[\boldsymbol{x}\right]_M \right\}  \right]^T$.
The operator $R(\cdot)$ converts a complex-valued vector into the equivalent real-valued notation. For a given column vector $\boldsymbol{a} \in \mathbb{C}^M$ the equivalent real-valued vector $\boldsymbol{a}_\text{r}=R(\boldsymbol{a})$ reads as \looseness-1
\begin{align}
\label{eq:stacked_vector_notation_for_mmse}
\boldsymbol{a}&_{\text{r}}=		\begin{bmatrix} \mathrm{Re} \left\{\boldsymbol{a}_1\right\} \   \mathrm{Im} \left\{\boldsymbol{a}_1\right\} \
\cdots \
\mathrm{Re} \left\{\boldsymbol{a}_M\right\} \  \mathrm{Im} \left\{\boldsymbol{a}_M\right\}
\end{bmatrix}^T.
\end{align}
The inverse operation is denoted as $C(\cdot)$ which converts equivalent real-valued notation into complex-valued notation.

%% file: 02_System_Model.tex
\section{System Model}
\label{sec:system_model}
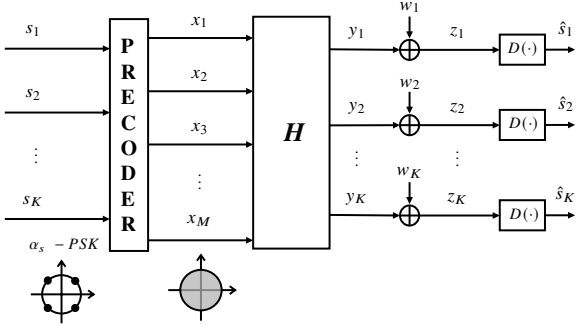
\begin{figure}
\captionsetup{justification=centering}
\centering
\input{figures/system_model}
\caption{Multiuser MIMO downlink with phase quantization and hard detection}
\label{fig:system_model}       
\end{figure}
The system model, illustrated in Fig.~\ref{fig:system_model}, consists of a single-cell MU-MIMO scenario where the BS has perfect channel state information (CSI) and is equipped with $M$ transmit antennas serving $K$ single antenna users.

A symbol level transmission is considered where $s_k$ represents the data symbol to be delivered for the $k$-th user. Each symbol $s_k$ is considered to belong to the set  $\mathcal{S}$ that represents all possible symbols of a $\alpha_{s}$-PSK modulation and is given by 
\begin{align}
	    \mathcal{S}= \left\{s: s= e^  \frac{j\pi (2 i+1) }{\alpha_{s}}  \textrm{,  for  }  i=1,\ldots, \alpha_{s} \right\}  \textrm{.}
	    \label{S_set}
\end{align}
The symbols of all users are described in a stacked vector notation as $\boldsymbol{s}=[{s}_1,\ldots,{s}_K]^T \in \mathcal{S}^K$. Based on $\boldsymbol{s}$ the precoder computes the transmit vector $\boldsymbol{x}=[x_{1} ,\hdots, x_{M}]^{T}$. The entries of $\boldsymbol{x}$ are constrained to the set $\mathcal{X}$ which represents a SPAPC and is given by
\begin{align}
	    \mathcal{X}= \chav{x: \PM{x}^2 \leq {\text{P}_\text{A}}},
	    \label{X_set}
\end{align}
where $\text{P}_\text{A}$ represents the maximum per antenna transmit power for a given time slot.
The vector $\boldsymbol{x}$ is transmitted over a frequency flat fading channel described by the matrix $\boldsymbol{H} \in \mathbb{C}^{K\times M}$ such that the received signal corresponding to the $k$-th user reads as
\begin{align}
\label{eq:symbols_at_receiver}
    {z_k} &= y_k+{w_k} = \boldsymbol{h}_k\ \boldsymbol{x}+{w_k}\text{,}
\end{align}
where $y_k$ is the noiseless received signal from the $k$-th user, $\boldsymbol{h}_k$ is the $k$-th row of the channel matrix $\boldsymbol{H}$ and the complex random variable ${w_k}\sim \mathcal{CN}  ({0},\sigma_w^2)$ represents additive white Gaussian noise (AWGN).
Using stacked vector notation \eqref{eq:symbols_at_receiver} can be extended to 
\begin{align}
    \boldsymbol{z} &= \boldsymbol{y}+\boldsymbol{w} =\boldsymbol{H}\ \boldsymbol{x}+\boldsymbol{w}\text{,}
\end{align}
where $\boldsymbol{z}=\pr{z_{1} \hdots z_{K}}^{T}$, $\boldsymbol{y}=\pr{y_{1} \hdots y_{K}}^{T}$ and $\boldsymbol{w}=\pr{w_{1} \hdots w_{K}}^{T}$.
Each received symbol $z_k$ is, then, hard detected based on which decision region it belongs, meaning that $z_k$ is detected as $s_i$ if $z_k \in \mathcal{S}_i$. In the case of PSK modulation the decision regions are circle sectors with infinite radius and angular aperture of $2\theta$, where $\theta$ is given by $\theta=\nicefrac{\pi}{\alpha_s}$. The hard detection operation is denoted by the operator $D(\cdot)$, such that the detected symbol from the $k$-th user is given by $\hat{s}_k=D(z_k)$. Finally, the detected symbol vector is written as $\hat{\boldsymbol{s}}=\pr{\hat{s}_1, \hdots, \hat{s}_K}$.

%% file: figures/system_model.tex
\tikzset{every picture/.style={line width=0.75pt}} 

\begin{tikzpicture}[x=0.30pt,y=0.30pt,yscale=-1,xscale=1]

\draw   (306.46,75.44) -- (354.63,75.44) -- (354.63,368) -- (306.46,368) -- cycle ;
\draw    (174,112.76) -- (303.46,112.76) ;
\draw [shift={(306.46,112.76)}, rotate = 180] [fill={rgb, 255:red, 0; green, 0; blue, 0 }  ][line width=0.08]  [draw opacity=0] (8.93,-4.29) -- (0,0) -- (8.93,4.29) -- cycle    ;
\draw    (174,193.36) -- (303.46,193.36) ;
\draw [shift={(306.46,193.36)}, rotate = 180] [fill={rgb, 255:red, 0; green, 0; blue, 0 }  ][line width=0.08]  [draw opacity=0] (8.93,-4.29) -- (0,0) -- (8.93,4.29) -- cycle    ;
\draw    (174,327.7) -- (303.46,327.7) ;
\draw [shift={(306.46,327.7)}, rotate = 180] [fill={rgb, 255:red, 0; green, 0; blue, 0 }  ][line width=0.08]  [draw opacity=0] (8.93,-4.29) -- (0,0) -- (8.93,4.29) -- cycle    ;
\draw    (354.63,99.32) -- (484.09,99.32) ;
\draw [shift={(487.09,99.32)}, rotate = 180] [fill={rgb, 255:red, 0; green, 0; blue, 0 }  ][line width=0.08]  [draw opacity=0] (8.93,-4.29) -- (0,0) -- (8.93,4.29) -- cycle    ;
\draw    (354.63,166.49) -- (484.09,166.49) ;
\draw [shift={(487.09,166.49)}, rotate = 180] [fill={rgb, 255:red, 0; green, 0; blue, 0 }  ][line width=0.08]  [draw opacity=0] (8.93,-4.29) -- (0,0) -- (8.93,4.29) -- cycle    ;
\draw    (354.63,233.66) -- (484.09,233.66) ;
\draw [shift={(487.09,233.66)}, rotate = 180] [fill={rgb, 255:red, 0; green, 0; blue, 0 }  ][line width=0.08]  [draw opacity=0] (8.93,-4.29) -- (0,0) -- (8.93,4.29) -- cycle    ;
\draw    (354.63,354.57) -- (484.09,354.57) ;
\draw [shift={(487.09,354.57)}, rotate = 180] [fill={rgb, 255:red, 0; green, 0; blue, 0 }  ][line width=0.08]  [draw opacity=0] (8.93,-4.29) -- (0,0) -- (8.93,4.29) -- cycle    ;
\draw   (487.09,72.45) -- (583.42,72.45) -- (583.42,365.01) -- (487.09,365.01) -- cycle ;
\draw    (583.42,209.54) -- (669.25,209.28) ;
\draw [shift={(672.25,209.27)}, rotate = 179.83] [fill={rgb, 255:red, 0; green, 0; blue, 0 }  ][line width=0.08]  [draw opacity=0] (8.93,-4.29) -- (0,0) -- (8.93,4.29) -- cycle    ;
\draw   (672.25,209.27) .. controls (672.25,202.27) and (677.64,196.59) .. (684.29,196.59) .. controls (690.94,196.59) and (696.34,202.27) .. (696.34,209.27) .. controls (696.34,216.28) and (690.94,221.96) .. (684.29,221.96) .. controls (677.64,221.96) and (672.25,216.28) .. (672.25,209.27) -- cycle ; \draw   (672.25,209.27) -- (696.34,209.27) ; \draw   (684.29,196.59) -- (684.29,221.96) ;
\draw    (684.29,167.23) -- (684.29,191.84) ;
\draw [shift={(684.29,194.84)}, rotate = 270] [fill={rgb, 255:red, 0; green, 0; blue, 0 }  ][line width=0.08]  [draw opacity=0] (8.93,-4.29) -- (0,0) -- (8.93,4.29) -- cycle    ;
\draw    (696.34,209.27) -- (795,209.27) ;
\draw [shift={(798,209.27)}, rotate = 180] [fill={rgb, 255:red, 0; green, 0; blue, 0 }  ][line width=0.08]  [draw opacity=0] (8.93,-4.29) -- (0,0) -- (8.93,4.29) -- cycle    ;
\draw  (206.78,422.75) -- (286,422.75)(245.47,380.93) -- (245.47,459.29) (279,417.75) -- (286,422.75) -- (279,427.75) (240.47,387.93) -- (245.47,380.93) -- (250.47,387.93)  ;
\draw   (221.02,422.75) .. controls (221.02,409.4) and (231.97,398.57) .. (245.47,398.57) .. controls (258.98,398.57) and (269.92,409.4) .. (269.92,422.75) .. controls (269.92,436.11) and (258.98,446.94) .. (245.47,446.94) .. controls (231.97,446.94) and (221.02,436.11) .. (221.02,422.75) -- cycle ;
\draw  [color={rgb, 255:red, 0; green, 0; blue, 0 }  ,draw opacity=1 ][fill={rgb, 255:red, 0; green, 0; blue, 0 }  ,fill opacity=1 ] (259.94,403.06) .. controls (261.46,401.54) and (263.94,401.54) .. (265.47,403.06) .. controls (267,404.57) and (267,407.02) .. (265.47,408.53) .. controls (263.94,410.04) and (261.46,410.04) .. (259.94,408.53) .. controls (258.41,407.02) and (258.41,404.57) .. (259.94,403.06) -- cycle ;
\draw  [color={rgb, 255:red, 0; green, 0; blue, 0 }  ,draw opacity=1 ][fill={rgb, 255:red, 0; green, 0; blue, 0 }  ,fill opacity=1 ] (225.36,437.26) .. controls (226.88,435.75) and (229.36,435.75) .. (230.89,437.26) .. controls (232.42,438.77) and (232.42,441.22) .. (230.89,442.73) .. controls (229.36,444.24) and (226.88,444.24) .. (225.36,442.73) .. controls (223.83,441.22) and (223.83,438.77) .. (225.36,437.26) -- cycle ;
\draw  [color={rgb, 255:red, 0; green, 0; blue, 0 }  ,draw opacity=1 ][fill={rgb, 255:red, 0; green, 0; blue, 0 }  ,fill opacity=1 ] (225.36,402.51) .. controls (226.78,400.89) and (229.17,400.81) .. (230.7,402.32) .. controls (232.22,403.83) and (232.31,406.36) .. (230.89,407.98) .. controls (229.47,409.6) and (227.08,409.68) .. (225.55,408.17) .. controls (224.02,406.66) and (223.94,404.13) .. (225.36,402.51) -- cycle ;
\draw  [color={rgb, 255:red, 0; green, 0; blue, 0 }  ,draw opacity=1 ][fill={rgb, 255:red, 0; green, 0; blue, 0 }  ,fill opacity=1 ] (259.94,437.26) .. controls (261.46,435.75) and (263.94,435.75) .. (265.47,437.26) .. controls (267,438.77) and (267,441.22) .. (265.47,442.73) .. controls (263.94,444.24) and (261.46,444.24) .. (259.94,442.73) .. controls (258.41,441.22) and (258.41,438.77) .. (259.94,437.26) -- cycle ;

\draw  (380,417.42) -- (464.47,417.42)(421.26,373.26) -- (421.26,456) (457.47,412.42) -- (464.47,417.42) -- (457.47,422.42) (416.26,380.26) -- (421.26,373.26) -- (426.26,380.26)  ;
\draw  [fill={rgb, 255:red, 179; green, 179; blue, 179 }  ,fill opacity=0.73 ] (395.19,417.42) .. controls (395.19,403.31) and (406.86,391.88) .. (421.26,391.88) .. controls (435.66,391.88) and (447.33,403.31) .. (447.33,417.42) .. controls (447.33,431.52) and (435.66,442.96) .. (421.26,442.96) .. controls (406.86,442.96) and (395.19,431.52) .. (395.19,417.42) -- cycle ;
\draw   (798,190) -- (855,190) -- (855,228) -- (798,228) -- cycle ;
\draw    (583.42,113.54) -- (669.25,114.25) ;
\draw [shift={(672.25,114.27)}, rotate = 180.47] [fill={rgb, 255:red, 0; green, 0; blue, 0 }  ][line width=0.08]  [draw opacity=0] (8.93,-4.29) -- (0,0) -- (8.93,4.29) -- cycle    ;
\draw   (672.25,114.27) .. controls (672.25,107.27) and (677.64,101.59) .. (684.29,101.59) .. controls (690.94,101.59) and (696.34,107.27) .. (696.34,114.27) .. controls (696.34,121.28) and (690.94,126.96) .. (684.29,126.96) .. controls (677.64,126.96) and (672.25,121.28) .. (672.25,114.27) -- cycle ; \draw   (672.25,114.27) -- (696.34,114.27) ; \draw   (684.29,101.59) -- (684.29,126.96) ;
\draw    (684.29,72.23) -- (684.29,96.84) ;
\draw [shift={(684.29,99.84)}, rotate = 270] [fill={rgb, 255:red, 0; green, 0; blue, 0 }  ][line width=0.08]  [draw opacity=0] (8.93,-4.29) -- (0,0) -- (8.93,4.29) -- cycle    ;
\draw    (696.34,114.27) -- (795,114.27) ;
\draw [shift={(798,114.27)}, rotate = 180] [fill={rgb, 255:red, 0; green, 0; blue, 0 }  ][line width=0.08]  [draw opacity=0] (8.93,-4.29) -- (0,0) -- (8.93,4.29) -- cycle    ;
\draw   (798,95) -- (855,95) -- (855,133) -- (798,133) -- cycle ;
\draw    (583.42,322.54) -- (669.25,323.25) ;
\draw [shift={(672.25,323.27)}, rotate = 180.47] [fill={rgb, 255:red, 0; green, 0; blue, 0 }  ][line width=0.08]  [draw opacity=0] (8.93,-4.29) -- (0,0) -- (8.93,4.29) -- cycle    ;
\draw   (672.25,323.27) .. controls (672.25,316.27) and (677.64,310.59) .. (684.29,310.59) .. controls (690.94,310.59) and (696.34,316.27) .. (696.34,323.27) .. controls (696.34,330.28) and (690.94,335.96) .. (684.29,335.96) .. controls (677.64,335.96) and (672.25,330.28) .. (672.25,323.27) -- cycle ; \draw   (672.25,323.27) -- (696.34,323.27) ; \draw   (684.29,310.59) -- (684.29,335.96) ;
\draw    (684.29,281.23) -- (684.29,305.84) ;
\draw [shift={(684.29,308.84)}, rotate = 270] [fill={rgb, 255:red, 0; green, 0; blue, 0 }  ][line width=0.08]  [draw opacity=0] (8.93,-4.29) -- (0,0) -- (8.93,4.29) -- cycle    ;
\draw    (696.34,323.27) -- (795,323.27) ;
\draw [shift={(798,323.27)}, rotate = 180] [fill={rgb, 255:red, 0; green, 0; blue, 0 }  ][line width=0.08]  [draw opacity=0] (8.93,-4.29) -- (0,0) -- (8.93,4.29) -- cycle    ;
\draw   (798,304) -- (855,304) -- (855,342) -- (798,342) -- cycle ;
\draw    (855,114) -- (888,114) -- (890,114) ;
\draw [shift={(893,114)}, rotate = 180] [fill={rgb, 255:red, 0; green, 0; blue, 0 }  ][line width=0.08]  [draw opacity=0] (8.93,-4.29) -- (0,0) -- (8.93,4.29) -- cycle    ;
\draw    (855,209) -- (888,209) -- (890,209) ;
\draw [shift={(893,209)}, rotate = 180] [fill={rgb, 255:red, 0; green, 0; blue, 0 }  ][line width=0.08]  [draw opacity=0] (8.93,-4.29) -- (0,0) -- (8.93,4.29) -- cycle    ;
\draw    (855,323) -- (888,323) -- (890,323) ;
\draw [shift={(893,323)}, rotate = 180] [fill={rgb, 255:red, 0; green, 0; blue, 0 }  ][line width=0.08]  [draw opacity=0] (8.93,-4.29) -- (0,0) -- (8.93,4.29) -- cycle    ;

\draw (213.71,241.7) node  [scale=0.6] {$\vdots $};
\draw (420,275) node  [scale=0.6] {$\vdots $};
\draw (617.77,245.27) node  [scale=0.6] {$\vdots $};
\draw (745,245.27) node  [scale=0.6] {$\vdots $};
\draw (209.7,90.19) node  [scale=0.7] {$s_{1}$};
\draw (209.7,170.8) node  [scale=0.7] {$s_{2}$};
\draw (209.7,305.14) node  [scale=0.7] {$s_{K}$};
\draw (420,79.25) node  [scale=0.7] {$x_{1}$};
\draw (420,146.42) node  [scale=0.7] {$x_{2}$};
\draw (420,213.59) node  [scale=0.7] {$x_{3}$};
\draw (420,329.72) node  [scale=0.7] {$x_{M}$};
\draw (617.77,188.05) node  [scale=0.7] {$y_{2}$};
\draw (745,187.46) node  [scale=0.7] {$z_{2}$};
\draw (684.91,157.26) node  [scale=0.7]  {$w_{2}$};
\draw (330.54,221.72) node  [scale=0.8] [align=left] {\textbf{P}\\\textbf{R}\\\textbf{E}\\\textbf{C}\\\textbf{O}\\\textbf{D}\\\textbf{E}\\\textbf{R}};
\draw (539.51,218.56) node  [scale=1] [align=left] {\textit{\textbf{H}}};
\draw (249.09,361) node  [scale=0.6]  {$\alpha _{s} \ -PSK$};
\draw (617.77,93.05) node  [scale=0.7] {$y_{1}$};
\draw (745,92.46) node  [scale=0.7] {$z_{1}$};
\draw (684.91,60.26) node  [scale=0.7]  {$w_{1}$};
\draw (617.77,302.05) node  [scale=0.7] {$y_{K}$};
\draw (745,301.46) node  [scale=0.7] {$z_{K}$};
\draw (684.91,269.26) node  [scale=0.7]  {$w_{K}$};
\draw (827,112.5) node  [scale=0.6]  {$D( \cdot )$};
\draw (828,208) node  [scale=0.6]  {$D( \cdot )$};
\draw (829,322) node  [scale=0.6]  {$D( \cdot )$};
\draw (881.5,87.5) node  [scale=0.7] {$\hat{s}_{1}$};
\draw (881.5,182.5) node  [scale=0.7] {$\hat{s}_{2}$};
\draw (881.5,296.5) node  [scale=0.7] {$\hat{s}_{K}$};

\end{tikzpicture}

%% file: 03_State_of_the_art.tex
\section{State-of-the-art Symbol Level Precoding Design under a Strict Per Antenna Power Constraint}
\label{sec:state_of_the_art}
In this section, we revise different state-of-the-art SLP approaches developed under the SPAPC. The design objectives considered are ZF and MMDDT which are briefly explained in each subsection. 

\subsection{Zero Forcing Design}

The ZF criterion is based on the idea of eliminating the interference considered as harmful for detection. As proposed in \cite{Chen2020}, the SLP ZF under a SPAPC can be designed by imposing the ZF constraint
and scaling it to satisfy the SPAPC. With this the closed form solution for the ZF-SPAPC precoding matrix is given as follows
\begin{align}
    \boldsymbol{P}=\frac{\sqrt{\text{P}_\text{A}} \boldsymbol{H}^\dagger}{\displaystyle\max_{m\in \chav{1,\hdots,M}} \PM{\pr{\boldsymbol{H}^\dagger\boldsymbol{s}}}_m}
\end{align}
where $\boldsymbol{H}^\dagger$ is Moore Penrose pseudo-inverse of the matrix $\boldsymbol{H}$. After computing $\boldsymbol{P}$ the vector $\boldsymbol{x}$ is computed as $\boldsymbol{x}=\boldsymbol{P}\boldsymbol{s}$.

\subsection{MMDDT Design}

The MMDDT criterion \cite{CVX-CIO,MSM_precoder,Landau2017,General_MMDDT_BB,Chen2020} consists of maximizing the minimum distance to the decision's threshold, denoted by $\epsilon$, such that the noiseless received signal is as inside as possible of the decision region of the symbol of interest. In what follows $\epsilon$ is derived and an MMDDT precoding SPAPC optimization problem is formulated.

To determine $\epsilon$ the first step is to apply a rotation by $\mathrm{arg}\{s_i^*\}=-\phi_{s_i}$ to the coordinate system such that the symbol of interest is placed on the real axis. This can be done by multiplying both the symbol of interest $s_i$ and the noiseless received signal $z_i$ by $e^{-j\phi_{s_i}} = s_i^*$ which yields
\begin{equation}
\begin{aligned}
    \quad s_i^{'}=s_i s_i^*=1  \text{,}  \quad  \omega_i=y_i s_i^* \textrm{.}
    \label{wi and si'}
\end{aligned}
\end{equation}
The distance of the rotated symbol $\omega_i$ to the rotated decision threshold is then expressed as
\begin{align}
    \epsilon_i=\mathrm{Re} \left\{\omega_i\right\}\sin{\theta}-\lvert \mathrm{Im} \left\{\omega_i\right\}\rvert\cos{\theta}  \textrm{.}
    \label{epsiloni_definition}
\end{align}
With this, $\epsilon$ is defined as 
\begin{align}
    \epsilon&=\min_{i\in \chav{1,\hdots,K}} \epsilon_i\notag\\
    &=\min_{i\in \chav{1,\hdots,K}} \mathrm{Re} \left\{\omega_i\right\}\sin{\theta}-\lvert \mathrm{Im} \left\{\omega_i\right\}\rvert\cos{\theta}. 
\end{align}
By introducing a stacked vector notation for $\omega_i$, in terms of $\boldsymbol{\omega}  = \mathrm{diag({\boldsymbol{s}^*})}  \boldsymbol{H} \boldsymbol{x} = \boldsymbol{H}_{s^*}\boldsymbol{x}$,
one can write the equivalent optimization problem as
\begin{align}
\label{opt:complex_mmddt}
	 \begin{bmatrix}\boldsymbol{x}_{\textrm{opt}} \textrm{,}\ \epsilon_\textrm{\hspace{0.2em}opt} 
	\end{bmatrix}&=\arg\min_{\boldsymbol{x}, \epsilon}  -\epsilon \\   \textrm{s.t.} \ \ 
 &\mathrm{Re}\left\{\boldsymbol{H}_{s^*}\boldsymbol{x}\right\}\sin{\theta}- \mathrm{Im}\left\{\boldsymbol{H}_{s^*}\boldsymbol{x}\right\} \cos{\theta}\geq \epsilon\boldsymbol{1} \notag\\ 
  &\mathrm{Re}\left\{\boldsymbol{H}_{s^*}\boldsymbol{x}\right\}\sin{\theta}+ \mathrm{Im}\left\{\boldsymbol{H}_{s^*}\boldsymbol{x}\right\} \cos{\theta}\geq \epsilon\boldsymbol{1} \notag\\
  & \PM{x_m}^2 \leq \text{P}_\text{A}, \ \ i\in \chav{1,\hdots,M}.\notag
\end{align} 
The problem above can be reformulated as a real-valued optimization problem by introducing the new optimization variable $\boldsymbol{u}=\pr{\epsilon, \boldsymbol{x}_\text{r}^T}^T$, with $\boldsymbol{x}_\text{r}=R(\boldsymbol{x})$, and reformulating the problem accordingly. With this, the equivalent real-valued optimization problem reads as
\begin{align}
\label{opt:rv_mmddt}
\boldsymbol{u}_{\text{opt}}=& \arg\min_{\boldsymbol{u}}  \boldsymbol{a}^T \boldsymbol{u} \\
 \textrm{s.t. } &\boldsymbol{B} \boldsymbol{u} \leq  \boldsymbol{0},  \notag \\
&     \|\boldsymbol{C}_\text{m} \boldsymbol{u}\|_2\leq \sqrt{\text{P}_\text{A}} , \ \ \textrm{for } m \in \chav{1,\hdots,M} ,  \notag
\end{align}
where
\begin{align}
\label{eq:matrix_a}
\quad &\boldsymbol{a}=[-1,\boldsymbol{0}^T]^T \text{,}
\quad\boldsymbol{B}= 
\begin{bmatrix} \boldsymbol{1} , \boldsymbol{\Theta}_{\textrm{r}}     
\end{bmatrix},
\\[5pt]
\quad &\boldsymbol{\Theta}_{\textrm{r}} =      \mbox{$\begin{bmatrix} 
\boldsymbol{\gamma}_1^T,
\boldsymbol{\lambda_1}^T, 
\cdots,
\boldsymbol{\gamma}_{K}^T, 
\boldsymbol{\lambda}_{K}^T, 
\boldsymbol{\psi}_1^T,
\boldsymbol{\delta}_1^T,  
\cdots, 
\boldsymbol{\psi}_K^T, 
\boldsymbol{\delta}_K^T
\end{bmatrix}$} \text{,} \\
\label{eq:Bm_matrix}
\quad&\boldsymbol{C}_\text{m}   =  \begin{bmatrix} 
 0 & \boldsymbol{0} \\[5pt]
 \boldsymbol{0}^T &\boldsymbol{D}_\text{m}
\end{bmatrix}, \quad \boldsymbol{D}_\text{m}= \text{diag}\pc{\boldsymbol{d}_\text{m}},
\end{align}
where $\boldsymbol{d}_\text{m} \in \mathbb{R}^{2M \times 1}$ being a vector of zeros with ones at entries $2m-1$ and $2m$, and $\boldsymbol{\gamma}_k$, $\boldsymbol{\lambda}_{k}$, $\boldsymbol{\psi}_{k}$ and $\boldsymbol{\delta}_k$ are the $k$-th row of the matrices $\boldsymbol{\mathrm{\Gamma}}$, $\boldsymbol{\mathrm{{\Lambda}}}$, $\boldsymbol{\mathrm{{\Psi}}}$ and $\boldsymbol{\mathrm{{\Delta}}}$, which are given by 
\begin{equation}
    \label{entries_of_Hr}
    \begin{aligned}
   \boldsymbol{\mathrm{\Gamma}}  & = & \mathrm{Im} \left\{  \boldsymbol{H}_{s^*}  \right\}\cos(\theta) & -\mathrm{Re} \left\{  \boldsymbol{H}_{s^*}  \right\} \sin(\theta) 
\\
    \boldsymbol{\mathrm{{\Lambda}}} & = &\mathrm{Re} \left\{  \boldsymbol{H}_{s^*}  \right\}\cos(\theta) & +\mathrm{Im} \left\{   \boldsymbol{H}_{s^*}  \right\} \sin(\theta)  
\\
          \boldsymbol{\mathrm{{\Psi}}}   & =&-  \mathrm{Im} \left\{  \boldsymbol{H}_{s^*}  \right\}\cos(\theta) & - \mathrm{Re} \left\{  \boldsymbol{H}_{s^*} \right\} \sin(\theta)  
\\
    \boldsymbol{\mathrm{{\Delta}}} & = &  \mathrm{Im} \left\{  \boldsymbol{H}_{s^*}  \right\}\sin(\theta) & -\mathrm{Re} \left\{   \boldsymbol{H}_{s^*}  \right\} \cos(\theta) \text{.}
\end{aligned}
\end{equation}
The problem described in \eqref{opt:rv_mmddt} is a SOCP cf.\cite[Sec. 4.4.2]{Boyd_2004} and can be readily solved with IPM. The optimal solution can be converted back to complex valued notation by extracting $\boldsymbol{x}_\text{r}$ of $\boldsymbol{u}_\text{opt}$ and applying $\boldsymbol{x}=C(\boldsymbol{x}_\text{r})$. Although the objectives are developed differently the MMDDT optimization problem described in \eqref{opt:rv_mmddt} is equivalent to the optimization problem developed for the Non-Strict CI-based design from \cite{Chen2020}.

%% file: 04_Precoding_task.tex
\section{Proposed MMSE Precoding Design under a Strict Per Antenna Power Constraint}
\label{sec:precoding}

In this section, we propose a SLP design based on the MMSE objective under a SPAPC. The MMSE objective, similar as proposed in \cite{M_Joham_ZF}, can be utilized under a SPAPC with the following optimization problem  
\begin{align}
\label{LRA_MMSE_precoder}
& \min_{\boldsymbol{x},\beta}  \mathrm{E} \chav{ \lVert    \beta {\boldsymbol{z}}  -  \boldsymbol{s}    \rVert_2^2
}\\
& \text{s.t. }     
  \PM{x_m}^2 \leq \text{P}_\text{A}, \ \textrm{for } \ m \in \chav{1,\hdots,M},\ \beta \geq 0
  \text{.} 
  \notag
\end{align}
Note that, the real-valued factor $\beta$ represents a theoretical automatic gain control which is part of the established MMSE objective as proposed in \cite{M_Joham_ZF}. Since, in this study, PSK modulation is considered, knowledge of $\beta$ is not required for hard detection. The problem can be rewritten by substituting $\boldsymbol{z}$ in the objective which yields
\begin{align}
\label{opt:mmse_prob1}
\quad
& \min_{\boldsymbol{x}_{\text{}},\beta}  \mathrm{E} \{  \lVert  \beta\boldsymbol{H}_{\text{}}\boldsymbol{x}_{\text{}} + \beta\boldsymbol{w}  -  \boldsymbol{s}   \rVert_2^2   \} \\
& \text{s.t. }     
  \PM{x_m}^2 \leq \text{P}_\text{A}, \ \textrm{for } \ m \in \chav{1,\hdots,M},\ \beta \geq 0
  \text{.} 
  \notag
\end{align}
Note that $\boldsymbol{x}$ is a complex-valued vector and $\beta$ a real-valued scaling factor. Rewriting the problem in a real-valued notation yields
\begin{align}
\label{opt:mmse_prob2}
& \min_{\boldsymbol{x}_{\text{r}},\beta}  \mathrm{E} \{  \lVert  \beta\boldsymbol{H}_{\text{r}}\ \boldsymbol{x}_{\text{r}} +\beta\boldsymbol{w}_{\text{r}}  -  \boldsymbol{s}_{\text{r}}\rVert_2^2   \} \\
&\text{s.t. } \beta\geq0  \text{,} \ \ \  \chav{\pr{\boldsymbol{x}_{\text{r}}}_{2m-1}^2+\pr{\boldsymbol{x}_{\text{r}}}_{2m}^2 }   \leq \text{P}_\text{A}, \notag\\
&\ \ \ \ \textrm{for } m \in \chav{1,\hdots,M}     \textrm{,}  \notag 
\end{align}
where $\boldsymbol{w}_\text{r}=R(\boldsymbol{w})$, $\boldsymbol{s}_\text{r}=R(\boldsymbol{s})$, with the operator $R(\cdot)$ introduced in \eqref{eq:stacked_vector_notation_for_mmse} and 
\begin{align}
&\boldsymbol{H}_{\textrm{r}} = \notag\\
&\begin{bmatrix} 
\textrm{Re}\chav{{h}_{11}} &-\textrm{Im}\chav{{h}_{11}} 
                         &\cdots&
\textrm{Re}\chav{{h}_{1M}} \ &-\textrm{Im}\chav{{h}_{1M}} 
\\
\textrm{Im}\chav{{h}_{11}} \  &\textrm{Re}\chav{{h}_{11}} 
                         &\cdots&
\textrm{Im}\chav{{h}_{1M}} \  &\textrm{Re}\chav{{h}_{1M}} 
\\
\vdots                         &&\ddots& &\vdots
\\
\textrm{Re}\chav{{h}_{K1}} \ &-\textrm{Im}\chav{{h}_{K1}} 
                         &\cdots&
\textrm{Re}\chav{{h}_{KM}} \ &-\textrm{Im}\chav{{h}_{KM}} 
\\
\textrm{Im}\chav{{h}_{K1}} \ &\textrm{Re}\chav{{h}_{K1}} 
                         &\cdots&
\textrm{Im}\chav{{h}_{KM}} \ &\textrm{Re}\chav{{h}_{KM}} 
\\
\end{bmatrix}  \text{.} \notag
\end{align}
The problem from \eqref{opt:mmse_prob2} can be expressed as an equivalent problem with
\begin{align}
\label{opt:mmse_prob3}
& \min_{\boldsymbol{x}_{\text{r}},\beta}  \beta^2 \boldsymbol{x}_{\text{r}}^T \boldsymbol{H}_{\text{r}}^T \boldsymbol{H}_{\text{r}}\ \boldsymbol{x}_{\text{r}}  -2 \beta    \boldsymbol{x}_{\text{r}}^T \boldsymbol{H}_{\text{r}}^T \boldsymbol{s}_{\text{r}}+ \beta^2 \mathrm{E} \{ \boldsymbol{w}_{\text{r}}^T\boldsymbol{w}_{\text{r}}  \}  \\
&\text{s.t. } \beta\geq0  \text{,} \ \ \  \chav{\pr{\boldsymbol{x}_{\text{r}}}_{2m-1}^2+\pr{\boldsymbol{x}_{\text{r}}}_{2m}^2 }   \leq \text{P}_\text{A}, \notag\\
&\ \ \ \ \textrm{for } m \in \chav{1,\hdots,M}     \textrm{,}  \notag 
\end{align}
where the objective function relates to the MSE as follows
\begin{align}
\label{eq:mse_tilde}
    \tilde{\text{MSE}}\pc{\boldsymbol{x}_\text{r},{\beta}}&= \text{MSE}\pc{\boldsymbol{x}_\text{r},{\beta}}- \text{E}\chav{\boldsymbol{s}_\text{r}^T \boldsymbol{s}_\text{r}} \\
    &= \beta^2 \boldsymbol{x}_{\text{r}}^T \boldsymbol{H}_{\text{r}}^T \boldsymbol{H}_{\text{r}}\ \boldsymbol{x}_{\text{r}}  -2 \beta    \boldsymbol{x}_{\text{r}}^T \boldsymbol{H}_{\text{r}}^T \boldsymbol{s}_{\text{r}}+ \beta^2 \mathrm{E} \{ \boldsymbol{w}_{\text{r}}^T\boldsymbol{w}_{\text{r}}  \}.\notag
\end{align}
Since $\text{E}\chav{\boldsymbol{s}_\text{r}^T \boldsymbol{s}_\text{r}}$ is constant, it is not relevant for the optimization. If  $\beta\geq0$ would be constant, the objective would be a convex quadratically constrained quadratic program (QCQP), since $\boldsymbol{H}_{\text{r}}^T\boldsymbol{H}_{\text{r}} \in S_{+}^{2M}$, \cite[Sec.~4.4]{Boyd_2004}.
Yet, the objective is in general not jointly convex in $\beta$ and $\boldsymbol{x}_{\text{r}}$ \cite[Appendix]{MMSE_bb}.
Nevertheless, it can be rewritten as an equivalent convex function by substituting the optimization variable $\boldsymbol{x}_\text{r}$. In this context, we introduce a new optimization variable $\boldsymbol{x}_\text{s} = \beta \boldsymbol{x}_{\text{r}} $, similar as done in \cite{lopes2021discrete,Jacobsson_2017}. With this, the optimization problem described in \eqref{opt:mmse_prob3} can be rewritten as
\begin{align}
& \min_{\boldsymbol{x}_{\text{s}},\beta}    \boldsymbol{x}_{\text{s}}^T  \boldsymbol{H}_{\text{r}}^T \boldsymbol{H}_{\text{r}} \boldsymbol{x}_\text{s}  -2    \boldsymbol{x}_\text{s}^T \boldsymbol{H}_{\text{r}}^T \boldsymbol{s}_{\text{r}}+\beta^2 \mathrm{E} \{ \boldsymbol{w}_{\text{r}}^T\boldsymbol{w}_{\text{r}}  \}  \\
&\ \ \text{s.t. } \pc{\frac{\pr{\boldsymbol{x}_\text{s}}_{2m-1}}{\beta}}^2+\pc{\frac{\pr{\boldsymbol{x}_\text{s}}_{2m}}{\beta}}^2  \leq \text{P}_\text{A}, \notag\\
&\ \ \ \ \ \ \ \textrm{for } m \in \chav{1,\hdots,M} \textrm{,} \ \   \beta\geq0  \text{,} \notag 
\end{align}
The first $M$ constraints can be reorganized by multiplying both sides with $\beta^2$ which yields  
\begin{align}
& \min_{\boldsymbol{x}_\text{s},\beta}    \boldsymbol{x}_\text{s}^T  \boldsymbol{H}_{\text{r}}^T \boldsymbol{H}_{\text{r}} \boldsymbol{x}_\text{s}  -2    \boldsymbol{x}_\text{s}^T \boldsymbol{H}_{\text{r}}^T \boldsymbol{s}_{\text{r}}+\beta^2 \mathrm{E} \{ \boldsymbol{w}_{\text{r}}^T\boldsymbol{w}_{\text{r}}  \}  \\
&\text{s.t. }  \pc{\pr{\boldsymbol{x}_\text{s}}_{2m-1}}^2+\pc{\pr{\boldsymbol{x}_\text{s}}_{2m}}^2  \leq \beta^ 2\text{P}_{\text{A}}, \notag\\
&\ \ \ \ \textrm{for } m \in \chav{1,\hdots,M},  \ \ \beta\geq0  \text{.}  \notag 
\end{align}
The problem can be written in matrix form as 
\begin{align}
\label{opt:mapped_mmse_problem}
& \min_{\boldsymbol{v}} \ \boldsymbol{v}^T \boldsymbol{U} \boldsymbol{v}  +  \boldsymbol{p}^T \boldsymbol{v}  \\ 
&\hspace{3mm} \text{s.t. }     
 \|\boldsymbol{E}_\text{m} \boldsymbol{v}\|_2\leq \boldsymbol{g}^T \boldsymbol{v} , \ \ \textrm{for } m \in \chav{1,\hdots,M} , \notag\\
 &\hspace{9mm} \boldsymbol{a}^T \boldsymbol{v} \leq 0 \notag
\end{align}
where
\begin{align}
\label{eq:quantities_mmse} 
\boldsymbol{v}=\begin{bmatrix}
\beta\\
\boldsymbol{x}_\text{s}
\end{bmatrix},
\quad \boldsymbol{U}    &=    \begin{bmatrix} 
 \mathrm{E} \{ \boldsymbol{w}_{\text{r}}^T\boldsymbol{w}_{\text{r}}  \} & \boldsymbol{0} \\
 \boldsymbol{0}^T & \boldsymbol{H}_\text{r}^T\boldsymbol{H}_\text{r}
\end{bmatrix},
\quad \boldsymbol{p}=\begin{bmatrix}
0 \\ -2 \boldsymbol{H}_{\text{r}}^T \boldsymbol{s}_{\text{r}}
\end{bmatrix}, \notag\\
\quad \boldsymbol{E}_\text{m}   &=  \begin{bmatrix} 
0 & \boldsymbol{0} \\
 \boldsymbol{0}^T &\boldsymbol{D}_\text{m}
\end{bmatrix},
\quad \boldsymbol{g}=\begin{bmatrix}
\sqrt{\text{P}_{\text{A}}}\\ \boldsymbol{0}
\end{bmatrix}, 
\end{align}
and $\boldsymbol{a}$ and $\boldsymbol{D}_\text{m}$ are described in \eqref{eq:matrix_a} and \eqref{eq:Bm_matrix}, respectively. Note that, the problem described in \eqref{opt:mapped_mmse_problem} is convex. In what follows it transformed into a SOCP in standard form, which significantly facilitates implementation. By introducing the additional variable $t$, cf. \cite[Sec. 4.1.3]{Boyd_2004}, the problem can be written with quadratic constraints as 
\begin{align}
\label{opt:epigraph}
& \min_{t,\boldsymbol{v}}  \ \boldsymbol{p}^T \boldsymbol{v} + 2t+1  \\ 
&\hspace{3mm} \text{s.t. }     
 \|\boldsymbol{E}_\text{m} \boldsymbol{v}\|_2\leq \boldsymbol{g}^T \boldsymbol{v} , \ \ \textrm{for } m \in \chav{1,\hdots,M} , \notag\\
& \hspace{9mm} \boldsymbol{v}^T \boldsymbol{U} \boldsymbol{v}  \leq 2t+1\notag\\
 &\hspace{9mm} \boldsymbol{a}^T \boldsymbol{v} \leq 0 \notag.
\end{align}
Note that, since $\boldsymbol{U} \in S_{+}^{2M+1}$, it can be written as $\boldsymbol{U}=\boldsymbol{L}^T\boldsymbol{L}$, with $\boldsymbol{L}=\boldsymbol{U}^{\frac{1}{2}}$. By substituting $\boldsymbol{U}=\boldsymbol{L}^T\boldsymbol{L}$ and adding $t^2$ at both sides of the quadratic constraint the problem is rewritten as
\begin{align}
\label{opt:intermediate_mmse_papc}
& \min_{t,\boldsymbol{v}}  \ \boldsymbol{p}^T \boldsymbol{v} + 2t  \\ 
&\hspace{3mm} \text{s.t. }     
 \|\boldsymbol{E}_\text{m} \boldsymbol{v}\|_2\leq \boldsymbol{g}^T \boldsymbol{v} , \ \ \textrm{for } m \in \chav{1,\hdots,M} , \notag\\
& \hspace{9mm} \boldsymbol{u}^T \boldsymbol{L}^T\boldsymbol{L}\boldsymbol{u} +t^2 \leq (t+1)^2\notag\\
 &\hspace{9mm} \boldsymbol{a}^T \boldsymbol{v} \leq 0 \notag.
\end{align}
By using stacked vector notation in the form of the new optimization variable $\boldsymbol{u}= \pr{ \boldsymbol{v}^T, t}^T$ and taking the square root of the quadratic constraint the optimization problem can be rewritten as
\begin{align}
\label{opt:intermediate_mmse_papc2}
& \min_{\boldsymbol{u}}  \ \boldsymbol{h}^T \boldsymbol{u}  \\ 
&\hspace{3mm} \text{s.t. }     
 \|\boldsymbol{F}_\text{m} \boldsymbol{u}\|_2\leq \boldsymbol{l}^T \boldsymbol{u} , \ \ \textrm{for } m \in \chav{1,\hdots,M} , \notag\\
& \hspace{8mm} \PM{\PM{\boldsymbol{G}\boldsymbol{u}}}_2 \leq \PM{t+1}\notag\\
 &\hspace{8mm} \boldsymbol{l}^T \boldsymbol{u} \leq 0 \notag
\end{align}
where
\begin{align}
& \boldsymbol{h}=\begin{bmatrix}
\boldsymbol{p} \\ 2
\end{bmatrix},
\quad \boldsymbol{F}_\text{m}   =    \begin{bmatrix} 
\boldsymbol{E}_\text{m}  & \boldsymbol{0} \\
 \boldsymbol{0}^T &  0
\end{bmatrix},
\quad\boldsymbol{l}=\begin{bmatrix}
\boldsymbol{a} \\
0
\end{bmatrix},
\quad\boldsymbol{G}   =  \begin{bmatrix} 
 \boldsymbol{L} & \boldsymbol{0} \\
 \boldsymbol{0}^T & 1
\end{bmatrix}.\notag
\end{align}
Considering the quadratic constraint in \eqref{opt:epigraph} and that $\boldsymbol{U} \in S_{+}^{2M+1}$ we have that $t\geq -\frac{1}{2}$, which then leads to $t+1\geq\frac{1}{2}>0$. With this, the problem described in \eqref{opt:intermediate_mmse_papc2} can be written in the standard form of a SOCP by considering $\PM{t+1}=t+1$ and rewriting the equivalent constraint in terms of the optimization variable $\boldsymbol{u}$. This, then, leads to 
\begin{align}
\label{opt:mmse_papc_socp}
& \min_{\boldsymbol{u}}  \ \boldsymbol{h}^T \boldsymbol{u}  \\ 
&\hspace{3mm} \text{s.t. }     
 \|\boldsymbol{F}_\text{m} \boldsymbol{u}\|_2\leq \boldsymbol{l}^T \boldsymbol{u} , \ \ \textrm{for } m \in \chav{1,\hdots,M} , \notag\\
& \hspace{8mm} \PM{\PM{\boldsymbol{G}\boldsymbol{u}}}_2 \leq \boldsymbol{q}^T\boldsymbol{u}+1\notag\\
 &\hspace{8mm} \boldsymbol{l}^T \boldsymbol{u} \leq 0 \notag,
\end{align}
with $\boldsymbol{q}=\pr{\boldsymbol{0}^T, 1}^T$. The problem described in \eqref{opt:mmse_papc_socp} is a SOCP, cf. \cite[Sec. 4.4.2]{Boyd_2004}, and can be readily solved with IPM. The solution can be converted back to complex valued notation by extracting $\boldsymbol{x}_\text{s}$ and $\beta$ from $\boldsymbol{u}_\text{opt}$ and applying $\boldsymbol{x}=C\pc{\frac{\boldsymbol{x}_\text{s}}{\beta}}$.

\subsection{About the Complexity of the Proposed Designs}

As mentioned the MMSE optimization problem is a SOCP and thus can be solved via IPM. According to \cite{ipm_complexity}, the number of iterations of the primal-dual IPM can be upper bounded by $\sqrt{n} \log\pc{\nicefrac{n}{\epsilon_\text{tol}}}$ where $n$ is the number of variables and $\epsilon_\text{tol}$ is the predefined optimality tolerance. Note that, the complexity of the iterations is dominated by solving a linear system needed to compute the primal-dual search direction. With this, considering that the linear systems can be solved with complexity $\mathcal{O}\pc{n^3}$ via Gauss-Jordan elimination, the total complexity of the proposed approach can be upper bounded by $\mathcal{O}\pc{M^{3.5}\log\pc{\nicefrac{M}{\epsilon_\text{tol}}}}$. Note that, for large-scale MIMO scenarios the complexity is dominated by the polynomial part. With this, the proposed problem have similar complexity as common optimization based precoders present in the literature \cite{CVX-CIO,MSM_precoder,Chen2020}.

%% file: 05_Numerical_Results.tex
\section{Numerical Results}
\label{sec:numerical_results}
\begin{figure}[t]
\begin{center}
\input{figures/papc_precoding}
\caption{BER versus $\mathrm{SNR}$ for $K=15$, $M=15$, $\alpha_s=4$} 
\label{fig:papc_precoding}       
\end{center}
\end{figure}
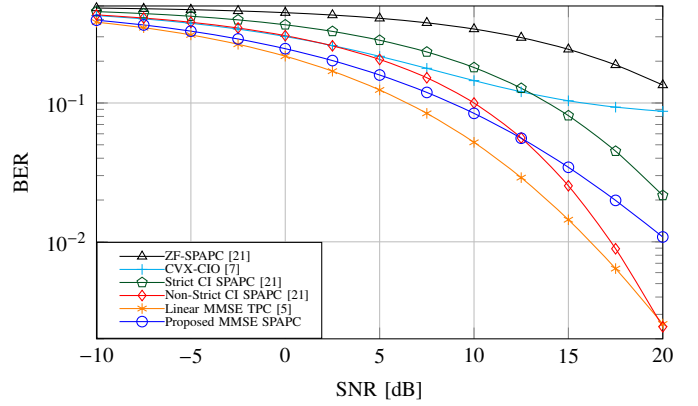
In this section, the proposed MMSE SPAPC precoder is evaluated in terms of BER and compared with other state-of-the-art designs. To this end, the SNR is defined as $\mathrm{SNR}= \nicefrac{\pc{M \cdot \text{P}_\text{A}}}{ \sigma_w^2}$. The channel coefficients are considered as Rayleigh fading \cite{Marzetta_2013} with large-scale fading coefficients equal to one.

\begin{figure}[t]
\begin{center}
\input{figures/papc_large}
\caption{BER versus $\mathrm{SNR}$ for $K=10$, $M=30$, $\alpha_s=8$} 
\label{fig:papc_large}       
\end{center}
\end{figure}
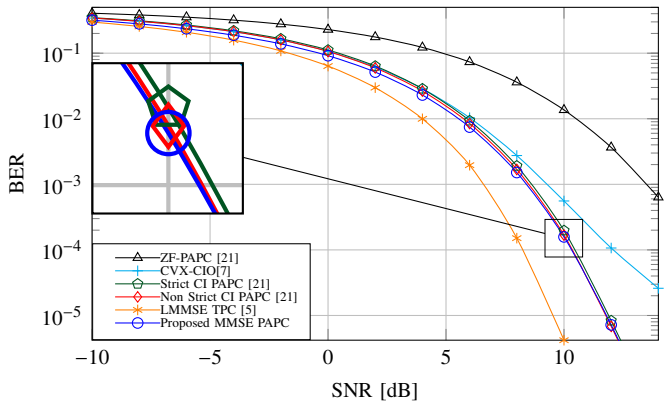

The proposed method is evaluated against the following state-of-the-art approaches: 1- The ZF SPAPC precoder \cite{Chen2020}; 2- The CVX-CIO precoder \cite{CVX-CIO} designed for constant envelope; 3- The Strict CI SPAPC precoder \cite{Chen2020}; 4- The Non-Strict CI SPAPC precoder \cite{Chen2020} and 5- The Linear MMSE precoder \cite{M_Joham_ZF} (average TPC).
\textcolor{black}{Two MIMO scenarios are evaluated. The first scenario consists in a BS with $M=15$ antennas serving $K=15$ users with user symbols drawn from a QPSK modulation, meaning that $\alpha_s=4$. The second considers a BS with $M=30$ antennas that serves $K=10$ users also with $\alpha_s=4$. The results are depicted in Fig.~\ref{fig:papc_precoding} and Fig.~\ref{fig:papc_large}. }

As can be seen in Fig.~\ref{fig:papc_precoding}, the proposed MMSE SPAPC method outperforms the existing approaches in terms of BER for the low and intermediate SNR regimes. For the high-SNR range the proposed MMSE SPAPC precoder outperforms all investigated approaches except of the Non-Strict CI based precoder which is equivalent to MMDDT design. This is expected since it is known that the MMSE criterion is favourable for low and intermediate SNR \cite{lopes2021discrete} while MMDDT is nearly optimal for the high SNR regime \cite{Jedda_2016}. Fig.~\ref{fig:papc_large} underlines the suitability of the proposed MMSE SPAPC approach for the intermediate SNR regime as it outperforms all state-of-the-art SPAPC approaches for this SNR range.

%% file: figures/papc_precoding.tex
%
%
%
\usetikzlibrary{positioning,calc}

\definecolor{mycolor1}{rgb}{0.00000,1.00000,1.00000}%
\definecolor{mycolor2}{rgb}{1.00000,0.00000,1.00000}%

\pgfplotsset{every axis label/.append style={font=\footnotesize},
every tick label/.append style={font=\footnotesize}
}

\begin{tikzpicture}[spy using outlines={rectangle,magnification=7,connect spies}] 

\begin{axis}[%
name=ber,
ymode=log,
width  = 0.85\columnwidth,
height = 0.5\columnwidth,
scale only axis,
xmin  = -10,
xmax  = 20,
xlabel= {SNR  [dB]},
xmajorgrids,
ymin=2e-3,
ymax=0.5,
ylabel={BER},
ymajorgrids,
legend entries={
                },
legend style={at={(0,0.33)},anchor=south west,draw=black,fill=white,legend cell align=left,font=\tiny}
]

\addlegendimage{solid,no marks,color=black,fill=gray!20,mark=square}


\addplot+[smooth,color=black,solid, every mark/.append style={solid, fill=gray!20},mark=triangle,
y filter/.code={\pgfmathparse{\pgfmathresult-0}\pgfmathresult}]
table[row sep=crcr]{%
-10	  0.483262760000000     \\
-7.5  0.477729155000000     \\
-5	  0.470339192500000     \\
-2.5  0.460508710000000     \\
0	  0.447463592500000     \\
2.5	  0.430216602500000     \\
5	  0.407579657500000     \\
7.5	  0.378260924166667      \\
10	  0.341162315833333        \\
12.5  0.295906267500000        \\
15	  0.243676230833333            \\
17.5  0.188049345000000            \\
20	  0.134724039166667              \\
};

\addplot+[smooth,color=cyan,solid, every mark/.append style={solid, fill=cyan!50},mark=+,
y filter/.code={\pgfmathparse{\pgfmathresult-0}\pgfmathresult}]
  table[row sep=crcr]{%
-10	   0.426511537500000    \\
-7.5   0.403657301666667    \\
-5	   0.375108470000000    \\
-2.5   0.340771270833333    \\
0	   0.301508253333333    \\
2.5	   0.259274152500000    \\
5	   0.216901566666667    \\
7.5	   0.177768970833333    \\
10	   0.144999795000000     \\
12.5   0.120403710833333     \\
15	   0.103751548333333     \\
17.5   0.0933430491666667     \\
20	   0.0871534383333333      \\
};

\addplot+[smooth,color=dark_green,solid, every mark/.append style={solid, fill=red!20},mark=pentagon,
y filter/.code={\pgfmathparse{\pgfmathresult-0}\pgfmathresult}]
  table[row sep=crcr]{%
-10	  0.455418416666667  \\
-7.5  0.440819964166667  \\
-5	  0.421739470833333  \\
-2.5  0.397142754166667  \\
0	  0.366059942500000  \\
2.5	  0.327993118333333  \\
5	  0.283248456666667  \\
7.5	  0.233161710833333      \\
10	  0.180108325000000      \\
12.5  0.127740094166667      \\
15	  0.0810507450000000      \\
17.5  0.0449692183333333      \\
20	  0.0215782300000000      \\
};

 \addplot+[smooth,color=orange,solid, every mark/.append style={solid, fill=blue!20},mark=asterisk,
 y filter/.code={\pgfmathparse{\pgfmathresult-0}\pgfmathresult}]
   table[row sep=crcr]{%
-10	    0.382323525000000  \\
-7.5    0.348837395000000  \\
-5	    0.309340181666667  \\
-2.5    0.264884091666667  \\
0	    0.217414770833333  \\
2.5	    0.169523879166667  \\
5	    0.124153453333333       \\
7.5	    0.0842068741666667       \\
10	    0.0520759900000000            \\
12.5    0.0289889516666667             \\
15	    0.0144221075000000             \\
17.5    0.00641253750000000      \\
20	    0.00255687666666667 \\
 };

 \addplot+[smooth,color=red ,solid, every mark/.append style={solid, fill=cyan!20},mark=diamond,
 y filter/.code={\pgfmathparse{\pgfmathresult-0}\pgfmathresult}]
   table[row sep=crcr]{%
-10	    0.431316820833333    \\
-7.5    0.409492677500000    \\
-5	    0.381677006666667    \\
-2.5    0.347098077500000    \\
0	    0.305655390000000    \\
2.5	    0.258086659166667    \\
5	    0.206001411666667    \\
7.5	    0.151952698333333    \\
10	    0.0999759975000000    \\
12.5    0.0559483000000000    \\
15	    0.0253110341666667        \\
17.5    0.00890963333333333        \\
20	    0.00244100000000000        \\  
 };
 
 \addplot+[smooth,color=blue ,solid, every mark/.append style={solid, fill=cyan!20},mark=o,
 y filter/.code={\pgfmathparse{\pgfmathresult-0}\pgfmathresult}]
   table[row sep=crcr]{%
-10	    0.395050162500000    \\
-7.5    0.364944492500000    \\
-5	    0.329369391666667    \\
-2.5    0.289262855000000    \\
0	    0.246202384166667    \\
2.5	    0.202133675833333    \\
5	    0.159092625833333    \\
7.5	    0.119151277500000    \\
10	    0.0842104891666667    \\
12.5    0.0557333208333333    \\
15	    0.0344231425000000        \\
17.5    0.0198787850000000        \\
20	    0.0108349225000000        \\  
 };

\addplot[smooth,color=black,solid,mark=triangle,
y filter/.code={\pgfmathparse{\pgfmathresult-0}\pgfmathresult}]
  table[row sep=crcr]{%
	1 2\\
};\label{plot:M15_zf}

\addplot[smooth,color=cyan,solid,mark=+,
y filter/.code={\pgfmathparse{\pgfmathresult-0}\pgfmathresult}]
  table[row sep=crcr]{%
	1 2\\
};\label{plot:M15_cvx-cio}

\addplot[smooth,color=dark_green,solid,mark=pentagon,
y filter/.code={\pgfmathparse{\pgfmathresult-0}\pgfmathresult}]
  table[row sep=crcr]{%
	1 2\\
};\label{plot:M15_strict}

\addplot[smooth,color=orange,solid,mark=asterisk,
y filter/.code={\pgfmathparse{\pgfmathresult-0}\pgfmathresult}]
  table[row sep=crcr]{%
	1 2\\
};\label{plot:M15_lmmse}

\addplot[smooth,color=red,solid,mark=diamond,
y filter/.code={\pgfmathparse{\pgfmathresult-0}\pgfmathresult}]
  table[row sep=crcr]{%
	1 2\\
};\label{plot:M15_nonstrict}

\addplot[smooth,color=blue,solid,mark=o,
y filter/.code={\pgfmathparse{\pgfmathresult-0}\pgfmathresult}]
  table[row sep=crcr]{%
	1 2\\
};\label{plot:M15_mmse}


\node [draw,fill=white,font=\tiny,anchor= south  west] at (axis cs: -10,2e-3) {
\setlength{\tabcolsep}{0.5mm}
\renewcommand{\arraystretch}{.8}
\begin{tabular}{l}

\ref{plot:M15_zf}{ZF-SPAPC \cite{Chen2020}}\\
\ref{plot:M15_cvx-cio}{CVX-CIO} \cite{CVX-CIO}\\
\ref{plot:M15_strict}{Strict CI SPAPC} \cite{Chen2020}\\
\ref{plot:M15_nonstrict}{Non-Strict CI SPAPC} \cite{Chen2020}\\
\ref{plot:M15_lmmse}{Linear MMSE TPC \cite{M_Joham_ZF}}\\
\ref{plot:M15_mmse}{Proposed MMSE SPAPC}
\end{tabular}
};

\end{axis}

\end{tikzpicture}%

%% file: figures/papc_large.tex
%
%
%
\usetikzlibrary{positioning,calc}

\definecolor{mycolor1}{rgb}{0.00000,1.00000,1.00000}%
\definecolor{mycolor2}{rgb}{1.00000,0.00000,1.00000}%

\pgfplotsset{every axis label/.append style={font=\footnotesize},
every tick label/.append style={font=\footnotesize}
}

\begin{tikzpicture}[spy using outlines={rectangle,magnification=4,connect spies}] 

\begin{axis}[%
name=ber,
ymode=log,
width  = 0.85\columnwidth,
height = 0.5\columnwidth,
scale only axis,
xmin  = -10,
xmax  = 14,
xlabel= {SNR  [dB]},
xmajorgrids,
ymin=4.2e-6,
ymax=0.5,
ylabel={BER},
ymajorgrids,
legend entries={
                },
legend style={at={(0,0.33)},anchor=south west,draw=black,fill=white,legend cell align=left,font=\tiny}
]

\addlegendimage{solid,no marks,color=black,fill=gray!20,mark=square}


\addplot+[smooth,color=black,solid, every mark/.append style={solid, fill=gray!20},mark=triangle,
y filter/.code={\pgfmathparse{\pgfmathresult-0}\pgfmathresult}]
table[row sep=crcr]{%
-10    0.406835187500000            \\
-8     0.383248775000000            \\
-6     0.354238700000000            \\
-4     0.319175325000000            \\
-2     0.277463625000000            \\
 0     0.229269925000000            \\
 2     0.176398262500000            \\
 4     0.122479875000000            \\
 6     0.0734787625000000            \\
 8     0.0360826500000000            \\
10     0.0136461750000000            \\
12     0.00365876250000000            \\
14     0.000634900000000000            \\
};

\addplot+[smooth,color=cyan,solid, every mark/.append style={solid, fill=cyan!50},mark=+,
y filter/.code={\pgfmathparse{\pgfmathresult-0}\pgfmathresult}]
  table[row sep=crcr]{%
-10    0.342928787500000          \\
-8     0.305792287500000          \\
-6     0.262059050000000          \\
-4     0.212492050000000          \\
-2     0.159328500000000          \\
 0     0.106990075000000          \\
 2     0.0614555625000000          \\
 4     0.0287247000000000          \\
 6     0.0103605750000000          \\
 8     0.00274842500000000          \\
10     0.000557562500000000          \\
12     0.000106912500000000          \\
14     2.59875000000000e-05          \\
};

\addplot+[smooth,color=dark_green,solid, every mark/.append style={solid, fill=red!20},mark=pentagon,
y filter/.code={\pgfmathparse{\pgfmathresult-0}\pgfmathresult}]
  table[row sep=crcr]{%
-10    0.348474687500000           \\
-8     0.312230700000000           \\
-6     0.269208262500000           \\
-4     0.219739125000000           \\
-2     0.165918037500000           \\
 0     0.111789887500000           \\
 2     0.0636875500000000           \\
 4     0.0286795125000000           \\
 6     0.00934671250000000           \\
 8     0.00191067500000000           \\
10     0.000197412500000000           \\
12     8.33750000000000e-06           \\
14     2.50000000000000e-07           \\
};

\addplot+[smooth,color=red,solid, every mark/.append style={solid, fill=blue!20},mark=diamond,
y filter/.code={\pgfmathparse{\pgfmathresult-0}\pgfmathresult}]
  table[row sep=crcr]{%
-10   0.343693275000000             \\
-8    0.306643587500000             \\
-6    0.262888000000000             \\
-4    0.213042200000000             \\
-2    0.159382525000000             \\
 0    0.106179612500000             \\
 2    0.0596744625000000             \\
 4    0.0264587125000000             \\
 6    0.00846380000000000             \\
 8    0.00168738750000000             \\
10    0.000168350000000000             \\
12    6.83750000000000e-06             \\
14    2.25000000000000e-07             \\
};

 \addplot+[smooth,color=orange,solid, every mark/.append style={solid, fill=blue!20},mark=asterisk,
 y filter/.code={\pgfmathparse{\pgfmathresult-0}\pgfmathresult}]
   table[row sep=crcr]{%
-10     0.298401287500000          \\
-8      0.255785750000000          \\
-6      0.208410062500000          \\
-4      0.158157162500000          \\
-2      0.108296637500000          \\
 0      0.0636419500000000          \\
 2      0.0298024375000000          \\
 4      0.00997263750000000          \\
 6      0.00195952500000000          \\
 8      0.000151087500000000          \\
10      4.20000000000000e-06          \\
 };

 \addplot+[smooth,color=blue ,solid, every mark/.append style={solid, fill=cyan!20},mark=o,
 y filter/.code={\pgfmathparse{\pgfmathresult-0}\pgfmathresult}]
   table[row sep=crcr]{%
-10     0.318932037500000            \\
-8      0.279809212500000            \\
-6      0.235701675000000            \\
-4      0.187892937500000            \\
-2      0.138613487500000            \\
 0      0.0915302625000000            \\
 2      0.0514726375000000            \\
 4      0.0230145000000000            \\
 6      0.00746023750000000            \\
 8      0.00151355000000000            \\
10      0.000157925000000000            \\
12      7.18750000000000e-06            \\
14      2.00000000000000e-07            \\
 };

\addplot[smooth,color=black,solid,mark=triangle,
y filter/.code={\pgfmathparse{\pgfmathresult-0}\pgfmathresult}]
  table[row sep=crcr]{%
	1 2\\
};\label{plot:M30_zf}

\addplot[smooth,color=cyan,solid,mark=+,
y filter/.code={\pgfmathparse{\pgfmathresult-0}\pgfmathresult}]
  table[row sep=crcr]{%
	1 2\\
};\label{plot:M30_cvx-cio}

\addplot[smooth,color=dark_green,solid,mark=pentagon,
y filter/.code={\pgfmathparse{\pgfmathresult-0}\pgfmathresult}]
  table[row sep=crcr]{%
	1 2\\
};\label{plot:M30_strict}

\addplot[smooth,color=red,solid,mark=diamond,
y filter/.code={\pgfmathparse{\pgfmathresult-0}\pgfmathresult}]
  table[row sep=crcr]{%
	1 2\\
};\label{plot:M30_nonstrict}

\addplot[smooth,color=orange,solid,mark=asterisk,
y filter/.code={\pgfmathparse{\pgfmathresult-0}\pgfmathresult}]
  table[row sep=crcr]{%
	1 2\\
};\label{plot:M30_lmmse}

\addplot[smooth,color=blue,solid,mark=o,
y filter/.code={\pgfmathparse{\pgfmathresult-0}\pgfmathresult}]
  table[row sep=crcr]{%
	1 2\\
};\label{plot:M30_mmse}

\coordinate (spypoint) at (axis cs:10,1.5e-4);
\coordinate (spyviewer) at (axis cs:-6.79,5e-3);
\spy[width=2cm,height=2cm] on (spypoint) in node [fill=white] at (spyviewer);

\node [draw,fill=white,font=\tiny,anchor= south  west] at (axis cs: -10,4.2e-6) {
\setlength{\tabcolsep}{0.5mm}
\renewcommand{\arraystretch}{.8}
\begin{tabular}{l}

\ref{plot:M30_zf}{ZF-PAPC \cite{Chen2020}}\\
\ref{plot:M30_cvx-cio}{CVX-CIO}\cite{CVX-CIO}\\
\ref{plot:M30_strict}{Strict CI PAPC} \cite{Chen2020}\\
\ref{plot:M30_nonstrict}{Non Strict CI PAPC} \cite{Chen2020}\\
\ref{plot:M30_lmmse}{LMMSE TPC \cite{M_Joham_ZF}}\\
\ref{plot:M30_mmse}{Proposed MMSE PAPC}
\end{tabular}
};

\end{axis}

\end{tikzpicture}%

%% file: 06_Conclusions.tex
\section{Conclusions}
\label{sec:conclusions}

This study considers PSK modulation and a strict per antenna power constraint and proposes a symbol-level precoding approach based on the MMSE criterion. The proposed precoding design is formulated as a SOCP and is solved using the IPM with polynomial computational complexity. Numerical results confirm that the proposed design is superior to the existing techniques in terms of BER for the low and intermediate SNR regime.

%% file: 00_main.bbl
\begin{thebibliography}{10}
\providecommand{\url}[1]{#1}
\csname url@samestyle\endcsname
\providecommand{\newblock}{\relax}
\providecommand{\bibinfo}[2]{#2}
\providecommand{\BIBentrySTDinterwordspacing}{\spaceskip=0pt\relax}
\providecommand{\BIBentryALTinterwordstretchfactor}{4}
\providecommand{\BIBentryALTinterwordspacing}{\spaceskip=\fontdimen2\font plus
\BIBentryALTinterwordstretchfactor\fontdimen3\font minus
  \fontdimen4\font\relax}
\providecommand{\BIBforeignlanguage}[2]{{%
\expandafter\ifx\csname l@#1\endcsname\relax
\typeout{** WARNING: IEEEtran.bst: No hyphenation pattern has been}%
\typeout{** loaded for the language `#1'. Using the pattern for}%
\typeout{** the default language instead.}%
\else
\language=\csname l@#1\endcsname
\fi
#2}}
\providecommand{\BIBdecl}{\relax}
\BIBdecl

\bibitem{6G_Future_Directions}
L.~U. {Khan}, I.~{Yaqoob}, M.~{Imran}, Z.~{Han}, and C.~S. {Hong}, ``{6G
  Wireless Systems: A Vision, Architectural Elements, and Future Directions},''
  \emph{{{IEEE} Access}}, vol.~8, pp. 147\,029--147\,044, 2020.

\bibitem{Kammoun2014}
A.~Kammoun, A.~Müller, E.~Björnson, and M.~Debbah, ``{Linear Precoding Based
  on Polynomial Expansion: Large-Scale Multi-Cell MIMO Systems},'' \emph{IEEE
  Journal of Selected Topics in Signal Processing}, vol.~8, no.~5, pp.
  861--875, 2014.

\bibitem{Tenbrink_2013}
J.~Hoydis, S.~ten Brink, and M.~Debbah, ``{Massive MIMO in the UL/DL of
  Cellular Networks: How Many Antennas Do We Need?}'' \emph{IEEE Journal on
  Selected Areas in Communications}, vol.~31, no.~2, pp. 160--171, 2013.

\bibitem{Power_consumption}
F.~{Rusek}, D.~{Persson}, B.~K. {Lau}, E.~G. {Larsson}, T.~L. {Marzetta},
  O.~{Edfors}, and F.~{Tufvesson}, ``{Scaling Up MIMO: Opportunities and
  Challenges with Very Large Arrays},'' \emph{{{IEEE} Signal Process. Mag.}},
  vol.~30, no.~1, 2013.

\bibitem{M_Joham_ZF}
M.~{Joham}, W.~{Utschick}, and J.~A. {Nossek}, ``Linear transmit processing in
  {MIMO} communications systems,'' \emph{{{IEEE} Trans. Signal Process.}},
  vol.~53, no.~8, pp. 2700--2712, Aug 2005.

\bibitem{Swindlehurst2005}
C.~Peel, B.~Hochwald, and A.~Swindlehurst, ``{A vector-perturbation technique
  for near-capacity multiantenna multiuser communication-part I: channel
  inversion and regularization},'' \emph{IEEE Transactions on Communications},
  vol.~53, no.~1, pp. 195--202, 2005.

\bibitem{CVX-CIO}
P.~V. {Amadori} and C.~{Masouros}, ``Constant envelope precoding by
  interference exploitation in phase shift keying-modulated multiuser
  transmission,'' \emph{{{IEEE} Trans. Commun.}}, vol.~16, no.~1, pp. 538--550,
  Jan 2017.

\bibitem{WeiYu2007}
W.~Yu and T.~Lan, ``{Transmitter Optimization for the Multi-Antenna Downlink
  With Per-Antenna Power Constraints},'' \emph{IEEE Transactions on Signal
  Processing}, vol.~55, no.~6, pp. 2646--2660, 2007.

\bibitem{BoccardiSPAWC2006}
F.~Boccardi and H.~Huang, ``{Zero-Forcing Precoding for the MIMO Broadcast
  Channel under Per-Antenna Power Constraints},'' in \emph{2006 IEEE 7th
  Workshop on Signal Processing Advances in Wireless Communications}, 2006, pp.
  1--5.

\bibitem{zf_papc}
K.~Karakayali, R.~Yates, G.~Foschini, and R.~Valenzuela, ``{Optimum
  Zero-forcing Beamforming with Per-antenna Power Constraints},'' in \emph{2007
  IEEE International Symposium on Information Theory}, 2007, pp. 101--105.

\bibitem{MRT_papc}
C.~Feng and Y.~Jing, ``{Modified MRT and outage probability analysis for
  massive MIMO downlink under per-antenna power constraint},'' in \emph{2016
  IEEE 17th International Workshop on Signal Processing Advances in Wireless
  Communications (SPAWC)}, 2016, pp. 1--6.

\bibitem{lmmse_papc}
C.-E. Chen, ``{MSE-Based Precoder Designs for Transmitter-Preprocessing-Aided
  Spatial Modulation Under Per-Antenna Power Constraints},'' \emph{IEEE
  Transactions on Vehicular Technology}, vol.~66, no.~3, pp. 2879--2883, 2017.

\bibitem{Pi2012}
Z.~Pi, ``{Optimal transmitter beamforming with per-antenna power
  constraints},'' in \emph{2012 IEEE International Conference on Communications
  (ICC)}, 2012, pp. 3779--3784.

\bibitem{bjorn_globecom2016}
D.~Spano, M.~Alodeh, S.~Chatzinotas, and B.~Ottersten, ``{Per-Antenna Power
  Minimization in Symbol-Level Precoding},'' in \emph{2016 IEEE Global
  Communications Conference (GLOBECOM)}, 2016, pp. 1--6.

\bibitem{Bjorn_satellite}
D.~Spano, S.~Chatzinotas, J.~Krause, and B.~Ottersten, ``{Symbol-level
  precoding with per-antenna power constraints for the multi-beam satellite
  downlink},'' in \emph{2016 8th Advanced Satellite Multimedia Systems
  Conference and the 14th Signal Processing for Space Communications Workshop
  (ASMS/SPSC)}, 2016, pp. 1--8.

\bibitem{MSM_precoder}
H.~{Jedda}, A.~{Mezghani}, A.~L. {Swindlehurst}, and J.~A. {Nossek},
  ``Quantized constant envelope precoding with {PSK} and {QAM} signaling,''
  \emph{{{{IEEE} Trans. Wireless Commun.}}}, vol.~17, no.~12, pp. 8022--8034,
  Dec 2018.

\bibitem{masouros_twc2018}
A.~Li and C.~Masouros, ``{Interference Exploitation Precoding Made Practical:
  Optimal Closed-Form Solutions for PSK Modulations},'' \emph{IEEE Transactions
  on Wireless Communications}, vol.~17, no.~11, pp. 7661--7676, 2018.

\bibitem{manifold_ce}
F.~Liu, C.~Masouros, P.~V. Amadori, and H.~Sun, ``{An Efficient Manifold
  Algorithm for Constructive Interference Based Constant Envelope Precoding},''
  \emph{IEEE Signal Processing Letters}, vol.~24, no.~10, pp. 1542--1546, 2017.

\bibitem{Landau2017}
L.~T.~N. {Landau} and R.~C. {de Lamare}, ``Branch-and-bound precoding for
  multiuser {MIMO} systems with 1-bit quantization,'' \emph{{{IEEE} Wireless
  Commun. Lett.}}, vol.~6, no.~6, pp. 770--773, Dec 2017.

\bibitem{General_MMDDT_BB}
E.~S.~P. {Lopes} and L.~T.~N. {Landau}, ``{Optimal Precoding for Multiuser MIMO
  Systems With Phase Quantization and PSK Modulation via Branch-and-Bound},''
  \emph{{{IEEE} Wireless Commun. Lett.}}, vol.~9, no.~9, pp. 1393--1397, 2020.

\bibitem{Chen2020}
C.-E. Chen, ``{Computationally Efficient Constructive Interference Precoding
  for PSK Modulations Under Per-Antenna Power Constraint},'' \emph{IEEE
  Transactions on Vehicular Technology}, vol.~69, no.~8, pp. 9206--9211, 2020.

\bibitem{lopes2021discrete}
E.~S.~P. Lopes and L.~T.~N. Landau, ``{Discrete MMSE Precoding for Multiuser
  MIMO Systems with PSK Modulation},'' \emph{IEEE Transactions on Wireless
  Communications}, 2022.

\bibitem{Boyd_2004}
S.~Boyd and L.~Vandenberghe, \emph{Convex Optimization}.\hskip 1em plus 0.5em
  minus 0.4em\relax New York, NY, USA: Cambridge University Press, 2004.

\bibitem{MMSE_bb}
E.~S.~P. {Lopes} and L.~T.~N. {Landau}, ``{Optimal and Suboptimal MMSE
  Precoding for Multiuser MIMO Systems Using Constant Envelope Signals with
  Phase Quantization at the Transmitter and PSK Modulation},'' in \emph{WSA
  2020; 24th International ITG Workshop on Smart Antennas}, Hamburg, Germany,
  2020.

\bibitem{Jacobsson_2017}
S.~{Jacobsson}, G.~{Durisi}, M.~{Coldrey}, T.~{Goldstein}, and C.~{Studer},
  ``{Quantized Precoding for Massive MU-MIMO},'' \emph{{{IEEE} Trans.
  Commun.}}, vol.~65, no.~11, pp. 4670--4684, 2017.

\bibitem{ipm_complexity}
J.~Peng, C.~Roos, and T.~Terlaky, ``{New Complexity Analysis of the
  Primal—Dual Newton Method for Linear Optimization},'' \emph{Annals of
  Operations Research}, vol.~99, no.~1, pp. 23--39, 2000.

\bibitem{Marzetta_2013}
H.~{Yang} and T.~L. {Marzetta}, ``Performance of conjugate and zero-forcing
  beamforming in large-scale antenna systems,'' \emph{{{IEEE} J. Sel. Areas
  Commun.}}, vol.~31, no.~2, pp. 172--179, 2013.

\bibitem{Jedda_2016}
H.~Jedda, J.~A. Nossek, and A.~Mezghani, ``Minimum {BER} precoding in 1-bit
  massive {MIMO} systems,'' in \emph{Proc. of IEEE Sensor Array and
  Multichannel Signal Processing Workshop (SAM)}, Rio de Janeiro, Brazil, July
  2016.

\end{thebibliography}
